\documentclass[conference]{IEEEtran}
\IEEEoverridecommandlockouts
\usepackage{cite}
\usepackage{amsmath,amssymb,amsfonts}
\usepackage{algorithmic}
\usepackage{graphicx}
\usepackage{textcomp}
\usepackage{xcolor}
\usepackage{array}
\usepackage[caption=false,font=normalsize,
labelfont=sf,textfont=sf]{subfig}
\usepackage{textcomp}
\usepackage{stfloats}
\usepackage{url}
\usepackage{verbatim}
\usepackage{graphicx}
\usepackage{balance}
\def\BibTeX{{\rm B\kern-.05em{\sc i\kern-.025em b}\kern-.08em
    T\kern-.1667em\lower.7ex\hbox{E}\kern-.125emX}}
\begin{document}

\title{Internal Gravity Waves in the Earth's Ionosphere\\
\thanks{A. P. M.  thanks Science and Engineering Research Board (SERB), Govt. of India for support through a Core Research Grant with the sanction order no. CRG/2018/004475 dated 26 March 2019. D. C. acknowledges support from SERB for a national postdoctoral fellowship (NPDF) with sanction order no. PDF/2020/002209 dated 31 Dec 2020.  }
}
\author{  A. P. Misra  and
  Animesh Roy \\ Department of Mathematics, Siksha Bhavana,  
 Visva-Bharati (A Central University), Santiniketan-731 235, India \\
Email: apmisra@visva-bharati.ac.in; ORCID: 0000-0002-6167-8136 
 \and
    Debjani Chatterjee \\ Department of Applied Mathematics,   University of Calcutta,  Kolkata-700 009, India \\
Email: chatterjee.debjani10@gmail.com 
\and
 T. D. Kaladze \\  Vekua Institute of Applied Mathematics and E. Andronikashvili Institute of Physics, Tbilisi State University, Georgia \\
 Email: tamaz{\_}kaladze@yahoo.com;  ORCID: 0000-0002-4151-8792 }

\maketitle

\begin{abstract}
The theory  of  low-frequency internal gravity waves (IGWs) is readdressed in the stable stratified  weakly ionized Earth's ionosphere.  The formation of dipolar vortex structures and their dynamical evolution, as well as, the emergence of chaos in the wave-wave interactions are  studied both in presence and absence  of the Pedersen conductivity.   The latter is shown to inhibit the formation of solitary  vortices   and the onset of chaos.  
\end{abstract}


\section{Introduction}\label{sec-intro}
Internal gravity waves (IGWs) are omnipresent in the atmosphere and ocean, and are the low frequency counterparts of 
 known acoustic-gravity waves (AGWs). Such waves appear in the interior of a stratified conducting fluid and are stimulated by the buoyancy   providing the restoring force which opposes the vertical displacements of atmospheric particles under gravity. The typical horizontal scale for these waves range from   $10$  to $1000$ km and their intrinsic frequencies $(\omega)$ vary in  between the Coriolis parameter $f=2\Omega\sin\lambda$  (where $\Omega$ is the angular   velocity of the planet and $\lambda$   the latitude) and the  Brunt-V{\"a}is{\"a}l{\"a} frequency $\omega_g$  i.e., in the interval  $10^{-4}$ s$^{-1}<\omega<1.7\times\times10^{-2}$ s$^{-1}$ \cite{holton2012,fritts2003,swenson1995,picard1998,mitchell1998}. The IGWs play crucial roles in the transportation  of charged particles as well as transfer of momentum and energy as they propagate vertically  in the regions starting from the Earth's surface to the upper atmosphere and ionosphere. Furthermore, the IGWs can have relevance in  the forcast problem of earthquakes and  the generation of large scale zonal flows \cite{horton2008}, the formation of solitary vortices \cite{stenflo1987}, as well as chaos \cite{roy2019,pati2021}  and turbulence \cite{shaikh2008,chatterjee2021} in the Earth's atmosphere.    
 \par
A number of fascinating phenomena can occur when the Coriolis force due to the Earth's rotation  comes into the picture. It has been established that such force not only  manifests the   coupling of high frequency AGWs and the low frequency IGWs, but also significantly alters the resonance and cut-off frequencies \cite{kaladze2008a,kaladze2008b}.  Furthermore, in ionospheric layers the Amp{\'e}re force due to  the interaction of the induced ionospheric currents with the geomagnetic field can have strong influence on the dynamics of ionospheric charged particles. It has been shown that such force can lead to the wave damping and the damping rate is proportional to the Pedersen conductivity associated with it \cite{kaladze2008a,kaladze2008b}.   
\par 
Recently, a number of researchers have focused their attention to investigate the dynamics of  IGWs in different contexts. 
To mention a few, Bulatov et al.  \cite{bulatov2020} have shown that the IGWs can be generated  far from a non-local source of disturbances in a layer of stratified media of a finite depth.  The theory of  AGWs  was revisited by    Chatterjee et al. \cite{chatterjee2021} to study the nonlinear coupling of AGWs and IGWs by taking into account the effects of the Coriolis force both in the contexts of the Zakharov approach \cite{zakharov1972} and the wave kinetic theory approach \cite{mendonca2014,mendonca2015}.      Pati et al. \cite{pati2021} studied the multistable dynamics of nonlinear AGWs in ionospheric rotating  fluids by the influence of the Coriolis force. They have shown that the AGWs can evolve into periodic and chaotic states. Similar chaotic dynamics of AGWs but in a different context (without the effect of the Coriolis force) was reported  by  Roy  et al.  \cite{roy2019}.  On the other hand, the propagation characteristics   of electromagnetic IGWs have been studied by Kaladze  et al.  \cite{kaladze2018,kaladze2019}  in an ideally conducting incompressible medium in presence of a constant magnetic field.    It has been  shown that the   IGWs can couple even with Alfv{\'e}n waves. 
 \par 
In this paper, we consider the nonlinear  dynamics of IGWs in the weakly ionized Earth's ionosphere embedded in the constant geomagnetic field (at sufficiently high latitudes) and study the  evolution of solitary vortices with different space localization as well as the existence of chaos in a low dimensional model. The effects of the Amp{\'e}re force on the dynamical evolution of IGWs are demonstrated. It is shown that the Pedersen conductivity associated with the electrmagnetic force dissipates the wave energy thereby inhibiting the formation of a vortex structure with finite wave energy and the occurrence of chaos in the wave-wave interactions.     The manuscript is organized as follows. In Sec. \ref{sec-formulation}, the theoretical model along with the physical assumptions for the description of IGWs are discussed. The formation of dipolar vortex and the evolution of chaos are also demonstrated in two subsections \ref{sec-stationary-sol} and \ref{sec-chaos}.     Finally, results are concluded in Sec. \ref{sec-conclu}. 

\section{Theoretical formulation and nonlinear dynamics of IGWs} \label{sec-formulation}
We consider the nonlinear dynamics  and evolution of low-frequency IGWs in the Earth's ionosphere. Specifically, we focus on the dynamics of IGWs that are generated at high latitudes in the northern hemisphere in which the constant geomagnetic field is vertically downwards along the $z$-axis (increasing upward), i.e., $\mathbf{B}_0=-B_0\hat{z}$. The other two ($x$ and $y$) axes are assumed to be directed from the west to the east and from the south to the north directions. Furthermore, at high latitudes the Earth's angular velocity can have only a vertical component, i.e., $\mathbf{\Omega}=(0,0,\Omega_0)$. In the non-inductive approximation, one can consider only the current $\mathbf{j}$ that originates in the medium due to the flow of electrons and ions, and the action of the geomagnetic field on them imposes us to consider the Amp{\'e}re $(\mathbf{j}\times \mathbf{B}_0)$ force on the dynamics of charged particles.  Thus, typical forces that can influence the dynamics of ionospheric plasmas are namely, the pressure gradient force, the Amp{\'e}re force,   the Coriolis force and the gravitational force. However, depending on the regions of the Earth's atmosphere we consider, different forces can become dominant over the other(s).   For example,   in the $E$-layer with  $n/N\sim 10^{-8}-10^{-6},~B_0\sim 0.6\times 10^{-4}~\mathrm{T}$, $\nu_{ei}\sim\nu_{in}\sim10^3~\mathrm{s}^{-1},~\nu_{en}\sim10^4~\mathrm{s}^{-1}$ and $\sigma_H\sim3\times10^{-4}~\mathrm{S/m}$, both the Amp{\'e}re force and the Coriolis force can have significant impact on the charged particles, whereas  in the $D$ and $E$   layers, one can safely neglect the contributions of the  Amp{\'e}re force and  the Coriolis force respectively \cite{kaladze2008a,kaladze2008b}. Here, $n/N$ is the ratio between the equilibrium number densities of electrons or ions and neutrals, $B_0$ is the static magnetic field, and $\nu_{ei},~\nu_{en}$ and $\nu_{in}$, respectively, denote the effective electron-ion, electron-neutral and ion-neutral   collisional frequencies. 
\par  
As a starting point, we disregard any influence of the Amp{\'e}re force and the Coriolis force. 
Thus, in stratified weakly ionized quasi-neutral ionospheric plasma layers, the   dynamics of  charged particles can be described by the following continuity equation, momentum balance  equation and the equation of state.
\begin{equation}
 \frac{\partial \rho}{\partial t}+\nabla\cdot(\rho\mathbf{v})=0, \label{eq-cont}
 \end{equation}
\begin{equation}
   \rho\frac{d \mathbf{v}}{dt}=- \nabla p +  \rho\mathbf{g}, \label{eq-moment}
\end{equation}
 \begin{equation}
 \frac{d}{dt}\left(p-  c_s^2 \rho\right) =0, \label{eq-press}
\end{equation} 
where $d/dt=\partial/\partial t +\mathbf{v}\cdot\nabla$,  $\mathbf{v}$ is the fluid flow velocity in a uniform gravity field with acceleration $\mathbf{g}=(0,0,-g)$,  $p$   $(\rho)$ is the fluid pressure (mass density), and  $c_s$ is the acoustic speed. In quasi-neutral plasmas, we have  neglected  the inner electrostatic field and the vortex component of the self-generated electromagnetic field \cite{kaladze2008a,kaladze2008b}.
The background (unperturbed by waves) pressure $p_0$ and the mass density $\rho_0$,   stratified by the gravitational field,  are given by the hydrostatic equilibrium equation $dp_0(z)/dz=-\rho_0(z)g$ with $p_0(z)=c_s^2\rho_0(z)$ which gives   $\rho_0(z)=\rho_0(0)\exp(-z/H)$ and $p_0(z)=p_0(0)\exp(-z/H)$,  where   $H=c_s^2/g$ is the  density scale height of the   atmosphere.   
 \par 
 In what follows, we obtain a set of nonlinear equations for the evolution of IGWs. We note that the 
density variation $(\rho_1)$ by the IGWs scales as $\rho_1/\rho_0\sim (1-4)\times10^{-2}$, so that in the momentum equation \eqref{eq-moment} one can approximate $\rho d\mathbf{v}/dt$ as $\rho_0(z) d\mathbf{v}/dt$ (Boussinesq approximation). Furthermore, using the incompressibility condition $\nabla\cdot\mathbf{v}=0$ the   two-dimensional evolution equations for IGWs in the $xz$-plane with $\mathbf{v}=(u,0,w)$ can be obtained as \cite{stenflo1987,stenflo1995}  
 \begin{equation}
 \frac{\partial}{\partial t}\left(\Delta \psi -\frac{\psi}{4H^2} \right)+J(\psi, \Delta \psi) =-\frac{\partial \chi}{\partial x},\label{eq-psi} 
 \end{equation}
 \begin{equation}
 \frac{\partial \chi}{\partial t} +J(\psi,\chi)=\omega_g^2 \frac{\partial \psi}{\partial x}, \label{eq-chi}
 \end{equation}
where $\psi(x,z)$ is the velocity stream function, given by,
$u=-{\partial\psi}/{\partial z},~w={\partial\psi}/{\partial x}$;
$\chi(x,z)=g\rho_1/\rho_0(0)$ is the variable associated with the density perturbation $\rho_1$, $\omega_g=\sqrt{g/H}$ is the Brunt-V{\"a}is{\"a}l{\"a} (buoyancy) frequency  for the incompressible fluid,     $\Delta= {\partial^2}/{\partial x^2}+{\partial^2}/{\partial z^2}$ is the two-dimensional Laplacian,  and $J(a,b)=({\partial a}/{\partial x})({\partial b}/{\partial z})-({\partial a}/{\partial z})({\partial b}/{\partial x})$ is the Jacobian.
\par 
From \eqref{eq-psi} and \eqref{eq-chi} we note that the total wave energy is conserved, i.e., $\partial {\cal E}/\partial t=0$, implying that a steady state solution can exist,   
where  the energy    ${\cal E}$ is   given by  
 \begin{equation}
{\cal E}=\int\int\left[\frac{1}{2}\left(\nabla\psi\right)^2+\frac{1}{8}\frac{\psi^2}{H^2}+\frac{1}{2}\frac{\chi^2}{\omega_g^2}\right]dxdz. \label{eq-ener2}
\end{equation}
\par 
Before going into the nonlinear evolution of IGWs, it is pertinent to obtain the linear dispersion relation in the limit of small amplitude waves with frequency $\omega$ and the wave vector $\mathbf{k}=(k_x,k_z)$. Thus, we have 
\begin{equation}
 \omega^2=\frac{k_x^2\omega_g^2}{k^2+1/4H^2}, \label{eq-omega}
 \end{equation}
where $k^2=k_x^2+k_z^2$. From   \eqref{eq-omega} it can be assessed that 
 the  wave frequency varies in the regime  $10^{-4}$ s$^{-1}<\omega~(\lesssim\omega_g)<1.7\times10^{-2}$ s$^{-1}$ and the   wavelength  $k^{-1}\sim H\sim (5-10)$ Km.   Also,  
 the phase velocity $|v_\text{ph}|\equiv|\omega/k_x|\leq v_\text{ph}^\text{max}\sim2H\omega_g=2\sqrt{gH}$. Typically, at a reduced atmospheric height with $H\sim 6$ Km  so that $H\omega_g\sim250$ m/s, we have  $v_\text{ph}\sim v_g\sim v_\text{ph}^\text{max}\sim 500$ m/s, where $v_g$ denotes the group velocity. These estimates agree  with some existing observations    \cite{sindelarova2009}.    So, a source of charged particles moving   along the $x$-direction with a velocity larger than $v_\text{ph}^\text{max}$ can not resonantly interact  with the IGWs and hence no energy loss of the wave.    In this case, one can look for a stationary solution of IGWs, i.e.,  the localization of a pulse in a frame moving with a speed $|U|>v_\text{ph}^\text{max}$  along the $x$-axis. Such  nonlinear solitary  structures  are  supersonic and their amplitudes do  not decay with time due to the generation of the linear wave mode with   $|v_\text{ph}|\leq v_\text{ph}^\text{max}$.
 Furthermore, we are interested in the frequency regime which is much higher than that due to the Coriolis acceleration, i.e., $\partial/\partial t\gg f$, where    $f=2\Omega\sin\lambda \sim7\times10^{-5}~\mathrm{rad/s}$   with $\lambda$ denoting the latitude. 
 \par 
 Nevertheless, the appearance of the nonlinear Jacobian terms   in \eqref{eq-psi} and \eqref{eq-chi} open up many possibilities to investigate. For examples, they can give rise the formation of various coherent localized vortex structures \cite{shukla1998}  and the onset of chaos in the coupling  of different wave modes \cite{roy2019}. However, the scenario changes significantly when the influence of the Amp{\'e}re force enters the dynamics of IGWs.  In this case, the equation of motion \eqref{eq-moment} changes to
  \begin{equation}
   \rho\frac{d \mathbf{v}}{dt}=- \nabla p + \mathbf{j}\times\mathbf{B}_0+\rho\mathbf{g}, \label{eq-moment2}
\end{equation} 
 where the current $j$ is given by the reduced form of the generalized Ohm's law \cite{kaladze2008a}
\begin{equation}
\mathbf{j}=\sigma_\perp\mathbf{E}_{d\perp}+\frac{\sigma_H}{B_0}\mathbf{B}_0\times \mathbf{E}_d, \label{eq-j}
\end{equation}
in which   only  the dynamo field $\mathbf{E}_d=\mathbf{u}\times\mathbf{B}_0$ contributes to the electric field. Also, in \eqref{eq-j}, the suffix $\perp$ denotes the components perpendicular to the constant magnetic field; $\sigma_\perp\equiv\sigma_p$ and $\sigma_H$ are, respectively,  the perpendicular   (Pedersen)  and Hall conductivities. The inclusion of this new force leads to a modification of \eqref{eq-psi}, i.e., 
  \begin{align}
 \frac{\partial}{\partial t}\left(\Delta \psi -\frac{\psi}{4H^2} \right)&+J(\psi, \Delta \psi) =-\frac{\partial \chi}{\partial x}, \notag \\  
 &-\frac{\sigma_p B_0^2}{\rho_0}\left(\frac{\partial^2 \psi}{\partial z^2} -\frac{\psi}{4H^2} \right)\label{eq-psi2} 
 \end{align}
such that the coupled equations \eqref{eq-chi} and \eqref{eq-psi2}   describe the nonlinear dynamics of low-frequency IGWs with the  effects of the magnetic viscosity $(\eta_0=\sigma_pB_0^2/\rho_0)$. The latter, however, violates the conservation of wave energy since the energy equation changes to
\begin{equation}
  \frac{\partial {\cal E}}{\partial t}+\eta_0 \int\int \left[\left(\frac{\partial \psi}{\partial z}\right)^2+\frac{1}{4}\frac{\psi^2}{H^2} \right]dxdz=0, \label{eq-enr1}
  \end{equation}
i.e., $\partial {\cal E}/\partial t\leq0$ for $\eta_0>0$ implying that a steady state solution with finite wave energy does not exist, i.e., the IGW gets damped by the effect of $\eta_0$. The linear damping rate can be obtained as 
\begin{equation}
 \gamma=-\frac{1}{2}\eta_0\frac{k_z^2+1/4H^2}{k^2+1/4H^2}. \label{eq-gamma}
 \end{equation} 
From \eqref{eq-gamma} it is evident that the damping rate increases with increasing values of the magnetic field. For a set of parameters as in the Earth's $F$ layer, its value can be estimated as $|\gamma|\sim \eta_0\sim10^{-4}$ s$^{-1}$, i.e., the damping rate is relatively small compared to the IGW frequency $\omega\sim10^{-2}~\mathrm{s}^{-1}$. It follows that the linear IGWs can develop into a certain nonlinear stage (before it being dissipated) to form vortex structures whose evolution is governed by \eqref{eq-chi} and \eqref{eq-psi2}.   
In the following two subsections \ref{sec-stationary-sol} and \ref{sec-chaos} we will investigate \eqref{eq-psi}, \eqref{eq-chi}, and \eqref{eq-psi2} in more details, and  look for stationary vortex  solutions together with their dynamical evolution as well as the occurrence of chaos in low dimensional models.  
\subsection{Formation of dipolar vortex} \label{sec-stationary-sol}
  In a stationary frame $\xi=x-Ut$ and $z=z$ moving with a velocity $U$ along the $x$-axis, Shukla \cite {shukla1998} obtained from a set of equations similar to \eqref{eq-psi} and \eqref{eq-chi} with $\eta_0=0$ a class of solutions including chains of vortices and circular vortex with monotonic profile as well as dipolar vortex. Following \cite {shukla1998}  another undamped dipolar vortex solution  of \eqref{eq-psi}  and \eqref{eq-chi}, which is regular at the center and  which vanishes at  infinity,   can be obtained for $\eta_0=0$  by applying  the transformations  $\xi=x-Ut$ and $z=z$, and considering the   linear relation between  $\psi$ and $\chi$:  $ \chi=-(\omega_g^2/ U)\psi$ as
 \begin{equation}
 \psi(r,\phi)=aUF(r)\cos\phi, \label{eq-psi-sol}
 \end{equation}
where 
\begin{equation}
F(r)=\left\lbrace\begin{array}{cc}
\frac{p^2}{k^2}\frac{J_1(kr)}{J_1(ka)}-\frac{(p^2+k^2)r}{ak^2},& (r<a)\\
-\frac{K_1(pr)}{K_1(pa)}, & (r\ge a).
\end{array}\right.
\end{equation}                                                                                              
Here,  $\xi=r\sin\phi,~z=r\cos\phi$, $a$ is the vortex radius,    $p^2= (1/4H^2)-{\omega_g^2}/{U^2}>0$, and $J_n$   $(K_n)$ is the Bessel function of first kind   (the McDonald  function) of order $n$.
The  vortex solution \eqref{eq-psi-sol} is antisymmetric with respect to   $z$ and  the vorticity   $\zeta=-\Delta\psi$ is continuous at $r=a$. Also, since $p^2>0$, the vortex profile can propagate   along the $x$-axis with the supersonic velocity $|U|>v^{\text{max}}_\text{ph}$ without any resonance with the linear wave. 
The typical size of the vortex $a$ can be estimated   as
 \begin{equation}
 a\sim p^{-1}=\frac{2H|U|}{\left(U^2-4H^2\omega_g^2\right)^{1/2}}\approx 2H. \label{eq-esti-a}
\end{equation}                                                                                                                     
Thus, the vortex structure of IGWs (with size $a\sim 2H\sim 15$ Km), which is exponentially localized in space and is regular at the center,  can propagate   in east-west directions (i.e., along $x$-axis) with the supersonic velocity $U$ and inevitably preventing the generation of linear modes by the moving structure (as the linear wave speed is $v_\text{ph}\lesssim v^{\text{max}}_\text{ph}$).
 \par 
 Another class of vortex solution of Eqs. \eqref{eq-psi} and \eqref{eq-chi} can be obtained  by considering $ \chi=-\omega_g H\Delta\psi$. For the phase velocity $U<0$ one obtains 
\begin{equation}
\psi=AJ_1(kr)\cos\phi, \label{eq-psi-sol3}
\end{equation}
where $A$ is a constant and 
$k^2=(1/H^2)\left(1- {U}/{v^{\text{max}}_\text{ph}}\right)>0$.  
  We note  that the dipole vortex solution \eqref{eq-psi-sol3} is regular at the center   $r=0$  and its amplitude $(\propto1/\sqrt{kr})$  vanishes at infinity. The size of the vortex   can be estimated as $a\sim k^{-1}\sim H\sim 10$ Km. 
On the other hand, for large $U~(>0)$, we have the following  solution
\begin{equation}
\psi=BK_1(pr)\cos\phi, \label{eq-psi-sol4}
\end{equation}
where $B$ is a constant and 
$ p^2=(1/H^2)\left({U}/{4H\omega_g}-1\right)>0$ for $U>4H\omega_g\sim2v^{\text{max}}_\text{ph}$.
The solution \eqref{eq-psi-sol4} has the singularity at the center $r=0$ and it is exponentially $[\propto\exp(-pr)]$ localized in space.   Its size can be estimated as $a\sim p^{-1}\sim H\sim10$ Km.
\par 
 In what follows, we look for a vortex solution of the coupled system \eqref{eq-chi} and \eqref{eq-psi2} with a small effect of $\eta_0\sim o(\epsilon)$. To this end, we assume   $\chi=-\omega_g H\Delta\psi$ and    expand $\psi(t)$ with $t\sim {\cal O}(1/\eta_0)$ as  
$ \psi(t)=\psi^{(1)}(t,t_1)+\eta_0\psi^{(2)}(t,t_1)+{\cal O}(\eta_0^2)$. Thus, we obtain
\begin{align}
&\psi=A_0\exp(-\eta_0 t)J_1(kr)\cos\phi, \notag \\
&\chi=B_0\exp(-\eta_0 t)K_1(pr)\cos\phi, \label{eq-psi-sol5}
\end{align}
where $A_0$ and $B_0$ are constants. From \eqref{eq-psi-sol5} it is noted that the amplitude of the   vortex profile decays with time by the effects of the magnetic viscosity. However, such vortex structures survive for a time $t\sim1/\eta_0$, e.g.,   $t\sim 30-3\times10^6$ hrs for the $E$-layer and $30-300$ hrs for the $F$-layer before it disappears due to energy loss.
\par
\begin{figure}[htbp] 
\centerline{\includegraphics[width=3in,height=2.5in]{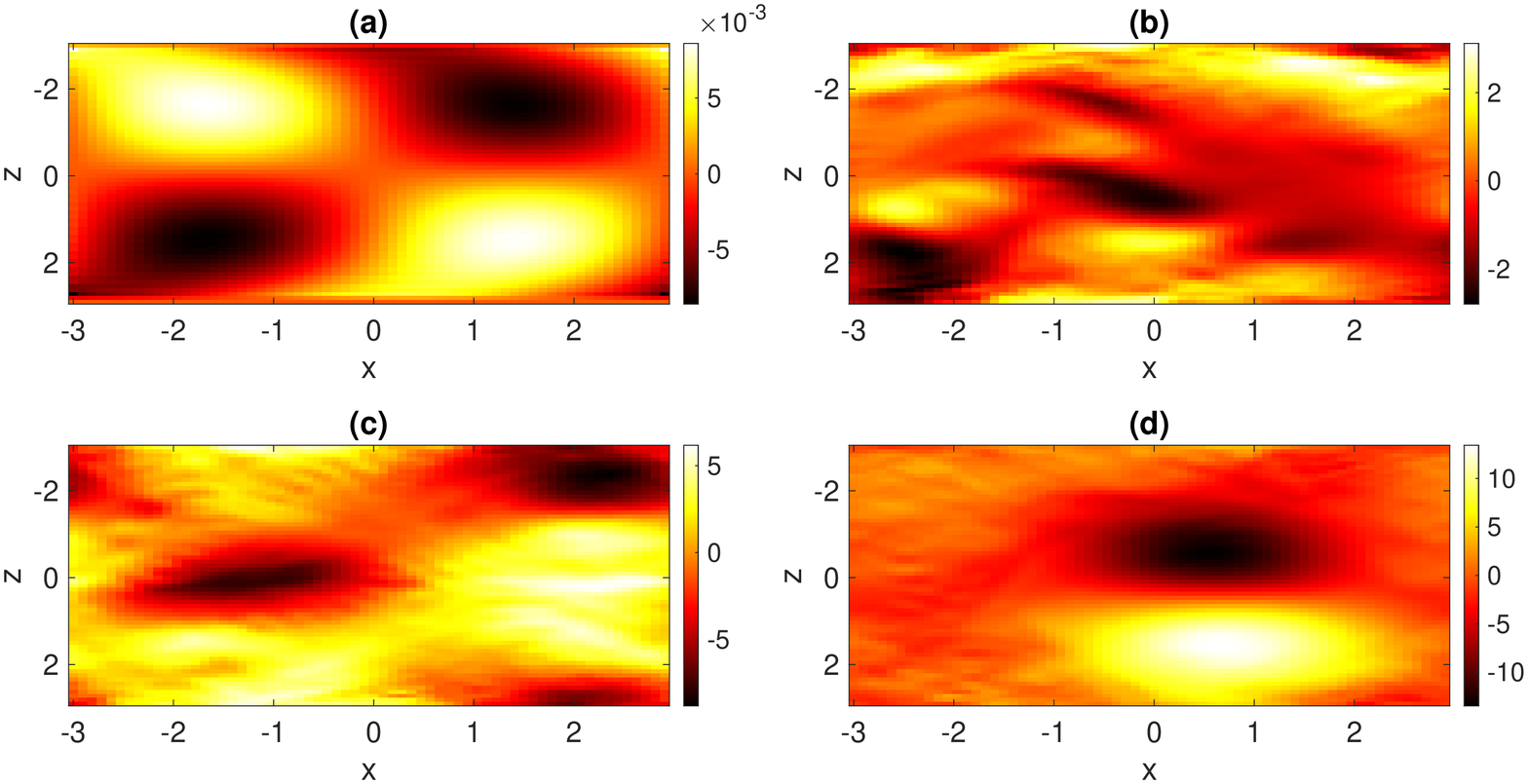}}
\caption{The evolution of solitary dipolar vortex [numerical solutions of \eqref{eq-psi} and \eqref{eq-chi}]  is shown for $\psi\sim\omega_g H^2$ at different times:  (a) $t=0$,    (b) $t=20\omega_g^{-1}$, (c) $t=40\omega_g^{-1}$,   and  (d) $t=70\omega_g^{-1}$. Here, $x,~z\sim H$ with $\omega_g\sim10^{-2}~\mathrm{s}^{-1}$  and  $H\sim10$ Km.}
\label{fig:vortex0}
\end{figure}

\begin{figure}[htbp] 
\begin{center}
\includegraphics[width=3in,height=2.5in]{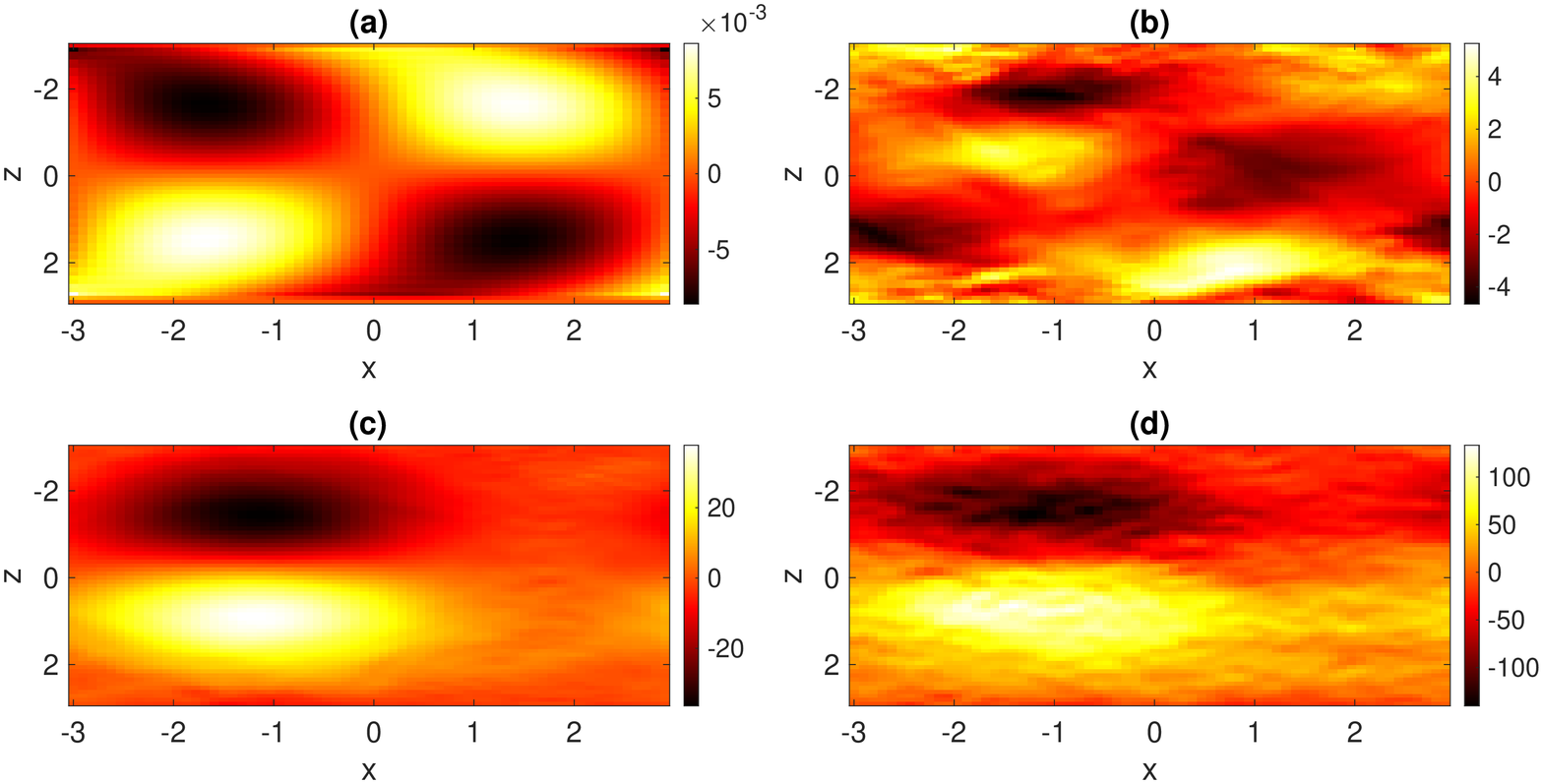}
\caption{The evolution of dipolar vortex [numerical solutions of \eqref{eq-chi} and \eqref{eq-psi2}]  with an effect of $\eta=0.2 \omega_g$ is shown for $\psi\sim\omega_g H^2$ at different times:  (a) $t=0$,    (b) $t=20\omega_g^{-1}$, (c) $t=50\omega_g^{-1}$,   and  (d) $t=70\omega_g^{-1}$. Here, $x,~z\sim H$ with $\omega_g\sim10^{-2}~\mathrm{s}^{-1}$  and  $H\sim10$ Km.    }
\label{fig:vortex1}
\end{center}
\end{figure}

 
In order to investigate the global behaviors of solitary vortices, we numerically solve the general system of equations \eqref{eq-chi} and \eqref{eq-psi2} by the $4$-th order Runge-Kutta scheme with a time step $dt\sim10^{-4}$ and mesh size $dx\sim dz\sim 0.1$. The evolution of dipolar vortex at different times with an initial profile being a two-dimensional sinusoidal wave, is shown in Figs. \ref{fig:vortex0} and \ref{fig:vortex1}. It is noted that in absence of the Pedersen conductivity $(\eta\equiv\eta_0/\omega_g=0)$, the solitary dipolar vortex is formed and it can propagate for a longer time without any distortion (see Fig. \ref{fig:vortex0}). However, as the effect of $\eta$  comes into the picture, the dipolar vortex, which was formed at $t=50\omega_g^{-1}$, tends to loose energy due to the dissipation and consequently, it completely  disappears after $t>70\omega_g^{-1}$ (see Fig. \ref{fig:vortex1}).
\subsection{Evolution of chaos} \label{sec-chaos}
\begin{figure*}[htbp] 
\centerline{\includegraphics[width=4.5in,height=2.5in]{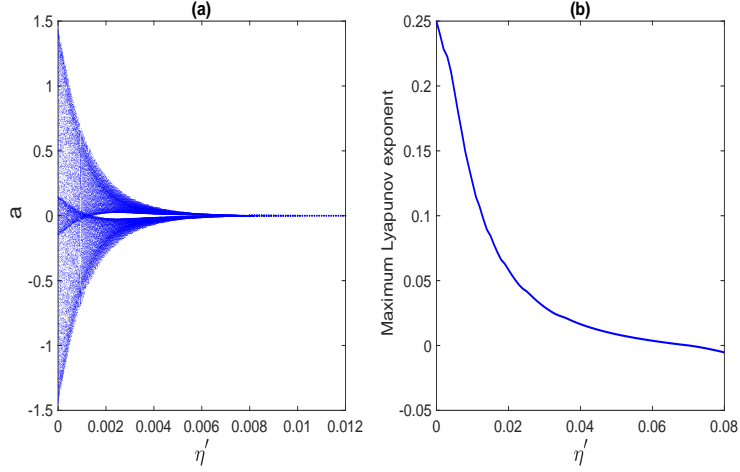}}
\caption{The bifurcation diagram [subplot (a)] and the maximum Lyapunov exponent [subplot (b)]  are shown with respect to the dissipative parameter $\eta^{\prime}=\eta/\left(1+1/4Hk_0^2\right)$ for the evolution of IGWs. The other parameter values are $\omega_0^{\prime}=0.3$ and $s=0.4$.   Here, $\eta^{\prime}=0$ correspond  to the system \eqref{eq-psi} and \eqref{eq-chi}, whereas $\eta^{\prime}\neq0$ corresponds to the dissipative  system \eqref{eq-chi} and \eqref{eq-psi2}. It is seen that the Pedersen conductivity $\eta$ has the effect of forbidding the occurrence of chaos in the nonlinear interaction of IGWs.  }

\label{fig:bifur-lyap}
\end{figure*}

\begin{figure*}[htbp] 
\centerline{\includegraphics[width=4in,height=3in]{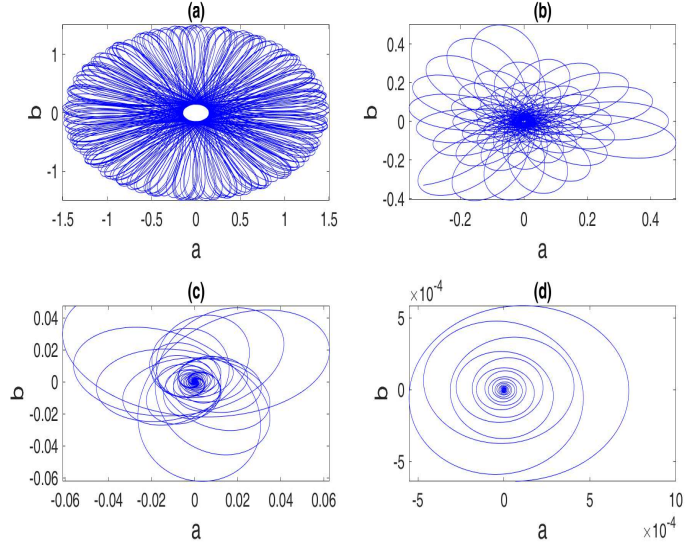}}
\caption{Trajectories flow of IGWs in phase space   is shown for different values of $\eta^{\prime}$: (a) $\eta^{\prime}=0$, (b) $\eta^{\prime}=0.01$, (c) $\eta^{\prime}=0.02$, and $\eta^{\prime}=0.03$. The other parameter values are $\omega_0^{\prime}=0.3$ and $s=0.4$. Here, $\eta^{\prime}=0$ correspond  to the system \eqref{eq-psi} and \eqref{eq-chi}, whereas $\eta^{\prime}\neq0$ corresponds to the dissipative  system \eqref{eq-chi} and \eqref{eq-psi2}.
Several asymptotic dynamics is seen where for a dissipative dynamical system volume  shrinks exponentially: (a)  chaotic strange attractor; (b) and (c) asymptotic fixed point dynamics via chaotic transit; (d) fixed point.  }
\label{fig:ph-space}
\end{figure*}

We note in  Sec. \ref{sec-stationary-sol} that the particular system \eqref{eq-psi} and \eqref{eq-chi}  or more generally, the system  \eqref{eq-chi} and \eqref{eq-psi2}, which  are multi-dimensional,  can admit localized dipolar vortex solutions with finite wave energy or  a solution whose amplitude decays due to dissipation. However, in the nonlinear interaction there may be a regime where a few coupled IGW modes are more active  than the remaining ones. In this case, a low-dimensional temporal model can describe some other important basic features of the full dynamics. Such cases are quite typical in the plasma wave turbulence \cite{banerjee2010}. 
Thus,  considering a class of solutions of    \eqref{eq-chi} and \eqref{eq-psi2} in the   form \cite{stenflo1987,roy2019}
\begin{align}
\psi&= \left[a(t)\sin(k_0x)+b(t)\cos(k_0x)+ \omega_0\right]z/k_0,\\
\chi& = \left[\alpha(t)\sin(k_0x)+\beta(t)\cos(k_0x)+\gamma(t)\right]z, \label{eq-ansatz}
\end{align}
where $k_0$ and $\omega_0$ are constants, we obtain the following  coupled dimensionless equations for  IGWs. 
\begin{align}
\dot{a}&=-\tilde{\omega}_0 b -s\beta-\eta^{\prime} a, \notag \\
\dot{b}& = \tilde{ \omega}_0 a + s\alpha-\eta^{\prime} b, \notag \\
\dot{\alpha}&=-\omega_0^{\prime}\beta+ b\gamma- b, \notag\\
\dot{\beta}&= \omega_0^{\prime}\alpha -a\gamma + a, \notag\\
\dot{\gamma}&=-a\beta + \alpha b.\label{eq-main}
\end{align}
Here, the overdot denotes differentiation with respect to $t$. Also, the variables $a$ and $b$ are normalized by $\omega_g$;  $\alpha$, $\beta$ and $\gamma$ are normalized by $\omega_g^2$. Furthermore, $\eta^{\prime}=s\eta$ and   $\tilde{\omega}_0 = s\omega_0^{\prime}$, where $\omega_0^{\prime}=\omega_0/\omega_g$ and $s=\left(1+1/4Hk_0^2\right)^{-1}$.  A careful numerical analysis of \eqref{eq-main} reveals that the system can undergo through periodic, quasiperiodic and chaotic states for different sets of parameter values \cite{roy2019}. Figure \ref{fig:bifur-lyap} shows the bifurcation diagram [subplot (a)] and the maximum Lyapunov exponent [subplot (b)] with respect to the parameter $\eta^{\prime}$ associated with the Pedersen conductivity. It is found that in absence of the Amp{\'e}re force (for which $\eta^{\prime}=0$), the system \eqref{eq-psi} and \eqref{eq-chi} exhibits chaos as evident from the positive Lyapunov exponent [subplot (b)]. However, the Lyapunov exponent tends to assume zero or negative values with increasing values of $\eta^{\prime}$.  As a result,  since the dissipation due to the Pedersen conductivity enters  the picture, several asymptotic dynamics is seen to occur where the flow volume  shrinks exponentially. Figure \ref{fig:ph-space} shows that a chaotic strange attractor forms   at  $\eta^{\prime}=0$. However, as its value increases,  the trajectories tend to move towards a fixed point  via chaotic transits.  The detailed investigation of the chaotic dynamics of \eqref{eq-chi} and \eqref{eq-psi2} is limited to the present work. Nevertheless, a detailed  dynamical study on a particular system \eqref{eq-psi} and \eqref{eq-chi} can be found in \cite{roy2019}. Here, one should note that some normalization issues occurred in the numerical analysis of \cite{roy2019}   can easily be fixed if one simply considers the values of $\omega_0$ as those of  $\omega_0/\omega_g$  and disregard the values of $\omega_g\sim1$, because   due to the normalization  of $a, ~b$ by $\omega_g$ and $\alpha,~\beta,~\gamma$ by $\omega_g^2$, the coefficient $\omega_g^2$ appeared in (6) of \cite{roy2019}  should disappear. 
 \section{Conclusion} \label{sec-conclu}
 We have studied the dynamical evolution of low frequency internal gravity waves (IGWs) in the weakly ionized Earth's ionosphere. Starting from a set of fluid equations relevant for incompressible conducting fluids, we have derived a set of coupled equations which are shown to admit localized dipolar solitary vortices. Both the analytical and numerical approach are performed to show that in absence of the Pedersen conductivity stationary dipolar vortex structures can be formed with finite wave energy which can propagate with supersonic velocity for a longer time and that IGWs can evolve into chaotic states due to wave-wave interactions in a low dimensional dimensional model. The latter, however, predicts some basic features of the full system and is useful in the context of wave turbulence in conducting fluids \cite{banerjee2010}.  As the dissipation due to the Pedersen conductivity enters the picture, the dipolar vortex structure tends to disappear due to the energy loss and the chaotic system  evolves towards a stable fixed point.    The results should be useful for understanding the remarkable features of low frequency internal gravity waves as well as the evolution of dipolar solitary vortices and chaotic flows associated with them in stratified weakly ionized plasmas that are relevant in the Earth's ionosphere.  The vortex structures so formed carry trapped particles and can play crucial roles in transportation of atmospheric components along the vertical direction thereby increasing the neutral density in the atmosphere at this height. Such density hike may   increase   the    recombination rate of  oxygen atoms and  hence   increase   the intensity of  night-sky radiation (with longer wavelengths  $\sim 550$ nm) that are observed before strong earthquakes. This enhancement of the radiation spectra can be useful for predicting earthquakes \cite{horton2008}.  
\section*{Acknowledgment}
A. P. M.  wishes to thank the organizer of the conference ICOPS2021 for   inviting him to prepare this manuscript.
\bibliographystyle{IEEEtran}
\bibliography{ReferenceIEEE.bib}{} 

\end{document}